\documentclass[%showpacs, 
twocolumn, preprintnumbers, amsmath,amssymb,nofootinbib]{revtex4}
\usepackage{graphicx}% Include figure files
\usepackage{dcolumn}% Align table columns on decimal point
\usepackage{bm}% bold math

\newcommand\ee{\end{equation}}
\newcommand\be{\begin{equation}}
\newcommand\eea{\end{eqnarray}}
\newcommand\bea{\begin{eqnarray}}
\newcommand{\sfrac}[2]{{\textstyle\frac{#1}{#2}}}
\newcommand\di{\partial}

\begin{document}

%\preprint{}

\title{A no-hair theorem for the galileon}% Force line breaks with \\

\author{Lam Hui}
\email{lhui@astro.columbia.edu}

\author{Alberto Nicolis}
\email{nicolis@phys.columbia.edu}

\affiliation{%
Physics Department and Institute for Strings, Cosmology, and Astroparticle Physics,\\
Columbia University, New York, NY 10027, USA
}%

\date{\today}% It is always \today, today,

\begin{abstract}
We consider a galileon field coupled to gravity. The standard no-hair
theorems do not apply, because of the galileon's peculiar
derivative interactions. 
We prove that, nonetheless, static spherically symmetric black holes
cannot sustain non-trivial galileon profiles.
Our theorem holds regardless of whether 
there are non-minimal couplings between the
galileon and gravity of the covariant galileon type.
%As a by-product, we show that in a two-dimensional space-time
%with a bifurcate Killing horizon, the isometry around the bifurcation
%point can always be expressed as Lorentz boost in suitable
%coordinates.
\end{abstract}

%\pacs{\dots}% PACS, the Physics and Astronomy
\maketitle

Black holes famously have no hair
\cite{bekenstein0,bekenstein1}---except when they
do \cite{EJW}.
No-hair theorems involve assumptions
that can be violated.
% In this paper, we are interested in scalar hair.
For instance, for scalar hair, Bekenstein's version 
\cite{bekenstein1} assumes that
the action depends on the derivatives of the scalar ($\pi$) only through
the combination $(\partial \pi)^2$. 
Several recent proposals for modifying gravity on long distance
scales involve introducing scalar derivative interactions of the galileon
type \cite{NRT}, which are not covered by existing no-hair theorems.
The galileon can even violate the null energy condition, in a
ghost-free manner \cite{NRT2}. In view of the observational and
theoretical significance of the galileon---observational
as an explanation for cosmic acceleration, theoretical as
a generic ingredient of massive gravity \cite{LPR, CGT}---it is useful to
investigate whether the no-hair theorem can be extended to
the galileon. We will demonstrate that indeed
black holes carry no galileon charge, at least for spherically
symmetric ones.
That this is true suggests an interesting experimental test of the
galileon, namely that central massive black holes in galaxies
are expected to be offset from the stars, which
is presented in a separate paper \cite{bhtest}.
A discussion of no-hair theorem for the galileon can also be
found in an independent paper by Babichev and Zahariade \cite{babichev}.
See also \cite{kaloper} for related discussions.

Our proof is logically very simple, and uses little information about the theory. 
The main ingredients it relies on are the shift-symmetry of the galileon action, the symmetries of the solution, and the regularity of diff-invariant quantities at the horizon.
It makes no use of Einstein's equations.
We assume that we have a spherically symmetric, static black-hole, in the presence of a spherically symmetric, static scalar field $\pi(r)$. We find it convenient to choose the radial coordinate $r$ such that $g_{tt} = -1/g_{rr}$, in which case the angular part of the metric has to be left generic:
\be	\label{metric}
ds^2 = - f(r) dt^2 + \frac1 {f(r)} dr^2 + \rho^2(r) d \Omega^2 \; .
\ee
Here $f$ and $\rho$ are generic functions. For the Schwarzschild
solution one has $f = 1-\frac{1}r$, $\rho=r$. For simplicity, we
choose units in which the black-hole horizon sits at $r=1$. We will
not assume the Schwarzschild metric, but rather use the more general
form eq. (\ref{metric}).
In the Appendix we collect a few technical statements about the
metric, perhaps of some general interest,
which follow from the regularity of curvature invariants at the
horizon. None of these statements are used in our proof of
the no-hair theorem, however. 
Our proof can be schematically divided into four steps:

\vspace{.2cm}

\noindent
{\em 1. The galileon eom is a current conservation equation.} The galileon equation of motion in the absence of sources can be written as the (covariant) conservation of a current:
\be \label{conservation}
\nabla_\mu J^\mu = 0 \; .
\ee
This follows directly from the shift invariance
\be
\pi \to \pi + c \; ,
\ee
which is exact for the galileon coupled to gravity, even in the presence of covariant galileon-type non-minimal couplings~\cite{DE-FV}.
The Noether current associated with such a symmetry involves first and second (covariant) derivatives of $\pi$---as well as curvature tensors in the covariant galileon case.
In flat space such a current takes the form \cite{NRT,EHHNW}
\be \label{J flat}
J^\mu =  G^{\mu\nu}(\di \di \pi) \, \di_\nu \pi \; ,
\ee
where $G$ is a tensor polynomial.

On the other hand, for generic spacetimes, galilean shifts of the form
\be
\pi \to \pi + b_\mu x^\mu \; 
\ee
are not a symmetry of the action, nor do they admit an obvious generalization
\footnote{See \cite{GHT, BdRH} for non-trivial generalizations to maximally symmetric spacetimes.}. 
As a consequence, their Noether currents \cite{N} are not conserved on non-trivial gravitational backgrounds. 
Our proof does not use them, and can thus be applied to generic shift-invariant scalar theories.

\vspace{.2cm}

\noindent
{\em 2. $J^r$ vanishes at the horizon.}
First, notice that in the coordinates that we are using, $J^r$ is the only non-zero component of $J^\mu$. This follows from the symmetries of the solution, for the metric as well as for the scalar. By rotational invariance, there cannot be any angular components of $J^\mu$. This is obvious, but here is a proof for the more fastidious among our readers. The current is a covariant quantity built out of the scalar and the metric, their derivatives, etc. The Killing vectors $\xi^\mu$ that generate rotations, define symmetries for the metric and for the scalar:
\be
{\cal L}_\xi \, g_{\mu\nu} = 0 \; , \qquad {\cal L}_\xi \, \pi = 0 \; ,
\ee
where ${\cal L}_\xi$ denotes the Lie-derivative along $\xi$. Any tensor constructed solely from these fields and their derivatives shares the same symmetries. In particular,
\be \label{Lie}
{\cal L}_\xi J^\mu = 0 \; .
\ee
However, we know that {\em any} regular two-vector field defined on a two-sphere must vanish at some point. Combining this with the spherical symmetry of $J^\mu$, eq.~\eqref{Lie}, we see that the angular components of $J^\mu$ have to vanish everywhere.

The vanishing of $J^t$ is slightly trickier to ascertain. From invariance under time-translations, following the same logic as above we just get that $J^t$ is constant in time, $\di_t J^t =0$. However, a non-zero $J^t$ picks out a time direction---it can be future-directed  or past-directed. There is nothing in the solution for the metric and for $\pi$ picking a time-direction---they only depend on $r$. And $J^t$ has to flip under time-reversal, because the galileon action does not contain an (odd number of) epsilon tensor(s): it only contains the scalar, the metric, and their derivatives, and it is invariant under time-reversal provided the scalar and the metric are even under it. We thus see that time-reversal invariance forces $J^t = 0$.

We are thus left with $J^r$ only. It is immediate to see that this has to vanish {\em at the horizon}.
As usual, the horizon $r=1$ corresponds to a zero of $f(r)$. This is because, by definition, the horizon corresponds to a locus where the time-translational Killing vector,
\be \label{xi}
\xi^\mu = (1,0,0,0) \; ,
\ee
becomes null: $g_{\mu\nu} \, \xi^\mu \xi^\nu = 0$.
Then, assuming that the horizon be a regular locus---a locus where all scalar quantities, physical and geometrical ones, are regular---we see that for $J^\mu J_\mu = (J^r)^2/f$ to be regular there, $J^r$ has to vanish.

\vspace{.2cm}

\noindent
{\em 3. $J^r$ vanishes everywhere.}  
We now use the current conservation equation \eqref{conservation} to bootstrap our way out of the near-horizon region, and show that, in fact, $J^\mu$ has to vanish everywhere.
Covariant current conservation can always be rewritten as
\be \label{spherical conservation}
\frac{1}{\sqrt{-g}}\di_\mu \big(  \sqrt{-g} J^\mu \big) = 0 \; .
\ee
In our case we have further simplifications. The only non-vanishing
component of $J^\mu$ is $J^r$, and it depends on $r$ only. Moreover,
in our coordinates $\sqrt{-g}$ is simply $\rho^2 (r) \sin^2 \theta$. 
Thus we have
$\rho^{-2} \partial_r (\rho^2 J^r) = 0$ which implies
\be
\rho^2 J^r = {\rm const} \; .
\ee
Notice that $\rho^2$ is expected to finite (neither infinite nor zero), even at the horizon, since it measures the area of constant-$r$ spheres.
We have shown previously that $J^r$ vanishes at the horizon,
and so the constant on the r.h.s.~is in fact zero. We therefore arrive at the conclusion
%
%We thus get
%\be \label{ODE}
%J^r {}' + 2 J^r \, \frac{\rho'}{\rho} = 0 \; ,
%\ee
%where the primes denote $r$-derivatives. This is a first-order ODE for $J^r(r)$, with---according to last item's result---boundary condition at the horizon $J^r(1) = 0$. The only subtlety in integrating this ODE is the behavior of the $\rho'/\rho$ factor at the horizon---is it regular or singular?
%
%To show that $\rho'/\rho$ is, in fact, regular, requires some work. We do this in the Appendix. The only assumptions used are: {\em (i)} the horizon has finite area; {\em (ii)} the horizon has finite surface gravity. Here by `finite' we mean `neither infinite nor zero'. The former assumption is obeyed by all horizons we know of. The latter is violated by extremal horizons. Taking for granted the regularity of $\rho'/\rho$, we have that  eq.~\eqref{ODE} is a regular, homogeneous, first-order ODE with vanishing boundary condition, whose only solution is 
\be \label{J is zero}
J^r = 0 \quad \mbox{at all $r$} \; . 
\ee

\vspace{.2cm}

\noindent
{\em 4. $\pi$ vanishes everywhere.}
The final step in our proof involves integrating eq.~\eqref{J is
  zero}, to find $\pi(r)$. Of course one possible solution is $\pi(r)
= 0$. We want to prove that this is in fact the only possible
solution. More precisely: it is the only solution that decays at
infinity
\footnote{Alternatively, one can say that we are interested in
a solution with vanishing first derivative at infinity, for $\pi =
{\,\rm constant}$ and $\pi = 0$ are equivalent solutions, related
by the shift symmetry.}.
To see this, note that for a spherically symmetric, static
configuration, the current takes the form
\begin{eqnarray} \label{Jr}
J^r = f \cdot  \pi' \cdot  F(\pi'; g, g',g'') \, ,
\end{eqnarray}
where $f = g^{rr}$ as in eq. (\ref{metric}), $\pi' \equiv d\pi/dr$, 
and $F$ is a polynomial of $\pi'$, whose coefficients depend on the metric and
its derivatives (to be justified below).
The crucial property of $F$ we will use is
that it asymptotes to a {\em nonzero constant} (which does not depend on the metric) when $\pi'$ goes to zero. This condition is obeyed
by any non-degenerate galileon theory featuring a kinetic energy for $\pi$. The reason is simply that in the weak $\pi$ limit, the action is well approximated by its quadratic terms and  the shift-current reduces simply to $J^\mu \simeq \di^\mu \pi$, up to an overall constant which defines $\pi$'s normalization.

Now, by assumption $\pi'$ vanishes at
infinity. Let us imagine dialing the radius to progressively smaller
values, starting from infinity. Imagine further that at some
radius, $\pi'$ starts deviating a little bit from zero. 
In that case, by continuity, $F$ will still be different from zero. Since $f$ does not vanish either (for $r > 1$),
we therefore reach the conclusion $J^r \ne 0$ (eq. \ref{Jr}), contradicting eq. (\ref{J is zero}). 
The resolution is that $\pi'$ in fact cannot deviate from zero,
thus $\pi' = 0$ at all radii, from which we conclude $\pi = {\rm
  const}$, or equivalently $\pi = 0$, completing our proof.

To round out our discussion, let us go back and justify the functional dependence of $F$.
The expression for the current can be derived straightforwardly from the action via Noether's theorem. 
For instance, $J^r$ has been computed explicitly for the galileon in flat space \cite{NRT}, where it takes the above form with $f =1 $ and 
\be \label{F flat}
F = F(\pi'/r) \qquad \mbox{(flat space).}
\ee
However the generic schematic form \eqref{Jr} can be inferred immediately by generalizing to general metric an argument given in \cite{NRT} for the flat-space case: For static, spherically symmetric configurations, the galileon equation of motion takes the form of the current conservation equation \eqref{spherical conservation}; Being a two-derivative eom, this cannot involve  derivatives of $\pi$ higher than $\pi''$; This then implies that $J^r$ cannot involve  derivatives of $\pi$ higher than $\pi'$; Moreover, $J^r$ cannot involve $\pi$ directly without derivatives, because each $\pi$ in the action is acted upon by at least one derivative. 
An analogous argument holds for the dependence of $F$ on the metric, and guarantees that $F$ depends at most on its  first $r$-derivatives for the covariant galileon case \cite{DE-FV}, and at most on its second $r$-derivatives for the minimally coupled one \cite{NRT}. 
We do not make use of this last fact---this is one of the reasons
why the covariant galileon and the minimal galileon cases can be
treated on equal footing in our proof.

%What we will make use of instead, is that as far $\pi$ is concerned, $F$ only depends on $\pi'$ and asymptotes to a {\em nonzero constant} (which does not depend on the metric) when $\pi'$ goes to zero. This condition is obeyed
%by any non-degenerate galileon theory featuring a kinetic energy for $\pi$. The reason is simply that in the weak $\pi$ limit, the action is well approximated by its quadratic terms and  the shift-current reduces simply to $J^\mu \simeq \di^\mu \pi$, up to an overall constant which defines $\pi$'s normalization.

%
%Indeed, far away from the black hole we can approximate the metric as flat, and consider the flat-space expression for the current, eq.~\eqref{J flat}. 
%For spherically symmetric, static configurations this can be rewritten as \cite{NRT}
%\be
%J^r \simeq \pi'(r) \, F\big( \pi'(r)/r \big ) \; , \qquad r \gg r_0 \; ,
%\ee
%where $F$ is a polynomial.
%The vanishing of $J^r$ requires either factor---$\pi'(r)$ or $F$---to vanish. However,
%if $\pi$ goes to zero at infinity, all non-linear terms in $J^r$ are negligible at very large---but finite---$r$. That is, the $F$ polynomial can be approximated by its zeroth order term, $F(0)$, which is nonzero for any non-degenerate galileon theory featuring a kinetic energy for $\pi$ \cite{NRT}. We thus have that $\pi'(r)$ has to vanish {\em exactly}---rather than to merely decay---at large $r$. This means that it has to vanish everywhere, and completes our proof.

\vspace{.2cm}

%{\bf Alberto: I am not sure how to extend the argument to the case
%where $\pi$ takes a profile such that what I call $1 + F$ above
%vanishes. There are 2 reasons why I am a bit confused.
%First, with non-minimal terms of the covariant galileon type, $F$
%depends on both $\pi$ and $\pi'$ (as well as the metric and its
%derivatives). With freedom to play with 2 quantities $\pi$ and $\pi'$,
%it seems one can't argue easily that if I pick a particular point in
%this 2-dimensional space where $1 + F = 0$, one can't find a small
%deformation that keeps $1 + F = 0$.
%Second, even if I ignore these non-minimal terms, so that $\pi'$
%is the only thing I can dial, it seems in general $\pi$ won't have
%exactly the $r^2$ profile, since the metric won't be exactly flat
%at finite $r$. Then, it becomes a question of what exactly is the
%nature of this ``tail'' that we are trying to rule out...
%Perhaps you have a better idea how to tackle this section.
%I have commented out the old version of this section in the tex file.
%}
%
\noindent
{\em Cosmological boundary conditions.}
For completeness, it is worth emphasizing that there can be in general non-linear solutions in which it is the $F$ factor that vanishes exactly, but these solutions will not decay at infinity---as our proof shows. Indeed, consider for the moment the flat-space case, where $F$ takes the form \eqref{F flat}. In order to make $\pi'/r$ constant and equal to a zero of $F$, these non-linear solutions have to behave as $\pi \sim r^2$ at large $r$, and correspond to non-trivial cosmological boundary conditions at infinity, rather than to scalar profiles ``sourced'' by the black-hole \cite{NR, NRT}.  Adding to this $r^2$ background field a ``tail'' that decays at infinity and that is somehow sourced by the black hole is not allowed, because that would violate the vanishing of the current. To see this, expand the full $\pi$ field as
\be \label{quadratic pi}
\pi(r) = \sfrac12 \beta \, r^2 + \delta \pi(r) \; ,
\ee
where $\beta$ is a zero of $F$, and $\delta \pi(r) $ is the decaying perturbation sourced by the black hole. At large $r$, we can expand eq.~\eqref{J is zero} in powers of $\delta \pi(r)$. The leading contribution comes from the first non-vanishing derivative of $F$, and yields simply
\be \label{delta pi}
\delta \pi'(r) = 0 \; ,
\ee
at large but {\em finite} $r$. 

This flat-space analysis is simplistic though, for it neglects two
sources of background curvature: the presence of the black-hole, and
the stress-energy tensor associated with $\pi$'s non-trivial
profile. In fact, even neglecting the latter, the former will
certainly affect $\pi$'s solution close to the black hole---the simple
$r^2$ profile will be distorted. 
Whether there is scalar hair now becomes a bit subtle---a matter of definition, to some extent.
In certain cases \cite{bhtest, HNS}, we are just interested in whether the black hole experiences
a $\pi$ force, i.e. whether it falls in the presence of some long
wavelength external $\pi$ fluctuation.
The question is thus whether far away from the black hole,
there is a $1/r$ tail to the $\pi$ profile on top of the $r^2$
cosmological background. Let us leave an explicit calculation to
the future, but instead consider the following alternative
characterization of `hair'.
%Whether there is scalar hair 
%now
%becomes---to some extent---a matter of definition: the $\pi \propto
%r^2$ profile will be modified close to the black hole, in a way that depends on the black hole under study: for bigger black-holes this distortion will start sooner, i.e.~at bigger $r$'s.
%So, in a sense, the black hole `sources' a $\pi$ field that corrects
%the background one. 
We can phrase the no-hair theorem as a statement of uniqueness: for given boundary conditions at infinity, static spherically symmetric black hole solutions are characterized just by one continuous parameter---the black-hole `mass'
\footnote{The quotes are needed because formally the black hole's ADM mass is not well defined for boundary conditions like ours, which curve spacetime at infinity.}.
That is, there is no scalar `charge' that can be varied independently of the mass. 
To see that this is indeed the case, consider for simplicity the limiting case of a black hole whose horizon radius $r_{\rm BH}$ is much much smaller than the curvature radius associated with the non-trivial (cosmological) boundary conditions---the Hubble radius $H^{-1}$. This is of course the relevant limit for astrophysical black holes in our universe. Consider an observer sitting at some intermediate radius 
\be
r_{\rm BH} \ll r \ll H^{-1} \; .
\ee
At such values of $r$, the metric can be well approximated by flat plus small corrections, the leading ones being a Schwartzschild-like tail and an FRW-like  quadratic potential \cite{NRT},
\be
g_{\mu\nu}(r) = \eta_{\mu\nu} + h_{\mu\nu} \;, \qquad h_{\mu\nu} = {\cal O} \big( \sfrac{2GM}{r}, H^2 r^2 \big) \; .
\ee
%\be
%ds^2 \simeq - \big[1-\sfrac{2GM}{r} \big] dt^2 + \big[ 1+ \sfrac{2GM}{r} \big] dr^2 + r^2 d\Omega^2 + {\cal O} (H^2 r^2) 
%\ee
%\be
%g_{\mu\nu} (r) \simeq g_{\mu\nu}^{M} (r) + {\cal O} (H^2 r^2) \; ,
%\ee
%where $g_{\mu\nu}^{M}$ is the Schwarzschild metric for a black hole of mass $M$; that is, in the same coordinates of eq.~\eqref{metric}, 
%\be
%f^{M}(r) = 1 - \sfrac{2GM}{r} \; ,  \qquad \rho^M(r) = r \; .
%\ee 
The only assumption here is that the total mass of the black hole---including any $\pi$ hair the black-hole might have---is finite, and that there is a radius $r$ much smaller than $H^{-1}$ that encloses most of this mass.
%The `cosmological' term of order $H^2 r^2$ can be made subleading by going close enough to the black-hole, 
%\be
%\frac{2GM}{r} \gg (H r)^2  
%\ee
Plugging this metric into $F(\pi'; g, g', g'') = 0$, and {\em algebraically} solving for $\pi'$, we get that $\pi'(r)$ depends on the same single parameter as the metric---the total mass $M$. In other words, the fact the $\pi$'s equation of motion reduces to an algebraic equation for the quantity we are directly interested in---$\pi'$---implies that there is no further integration constant involved in solving this equation. This is to be contrasted with, say, the case of electrically charged black holes: there, solving Maxwell's equations for the electric field entails introducing an integration constant---the total charge $Q$---which is independent of the black hole mass.
In this sense, our black holes carry no independent scalar hair.
%That is, the inclusion of non-trivial boundary conditions does not improve the situation---black hole-sourced galileon profiles are still forbidden.
%
%
%\footnote{Eq.~\eqref{delta pi} admits $\delta \pi= {\rm const} \neq 0$ as a solution. However this is not really a perturbation sourced by the black hole, but rather a different choice of boundary conditions, which, given the shift-symmetry of the galileon dynamics, is physically equivalent to the $\delta \pi = 0$ one.}.
%
%

\vspace{.2cm}

\noindent
{\em Concluding remarks.}
We have shown that static, spherically symmetric black-hole solutions
for the gravity-galileon coupled system cannot sustain non-trivial
galileon profiles,  for vanishing boundary condition at infinity.
Our proof does not make use of Einstein's equations---it only uses the
shift-symmetry of the galileon action, and 
the regularity of diff-invariant quantities at the horizon. 
In would be interesting to extend our analysis to stationary rotating black holes.

In the presence of non-trivial (cosmological) boundary conditions for
the galileon field, the question of scalar hair is more subtle. 
We have shown that for black holes that are much smaller than
the asymptotic curvature radius, the galileon solution at intermediate
scales cannot depend on `scalar charges'---i.e., integration
constants---other than the total mass of the black hole.
It would be useful to confirm the lack of scalar hair by explicitly showing that
the $\pi$ profile contains no $1/r$ tail. We leave this for future work.
It is also worth noting that known black hole solutions in
massive gravity are consistent with our results,
that they either have no galileon hair or suffer
from singularities at the horizon (see \cite{BCdGT} and references
therein).

\vspace{.3cm}
 
\noindent
{\em Acknowledgements.}
We would like to thank Gregory Gabadadze, Walter Goldberger, Nemanja Kaloper, Matt Kleban,
Federico Piazza, and Erick Weinberg for useful discussions and comments.
This work  is supported by the DOE under contracts DE-FG02-11ER41743 and DE-FG02-92-ER40699, and by NASA under contract NNX10AH14G.
 
\appendix

\section{Regularity at the horizon}
The horizon at $r = 1$ is a null surface with $\rho = {\rm const} \neq 0$, assuming a non-vanishing surface area. The normal to such a surface,
$\di_\mu \rho$, is thus null, that is, 
\be
f \rho'^2 \to 0 \qquad (r \to1) \; .
\ee

Consider the curvature invariant $R_{\mu\nu} R^{\mu\nu}$, which should
be regular (non-singular) at the horizon, and which for the metric \eqref{metric} reads
\begin{align}
R_{\mu\nu} R^{\mu\nu} = & \frac2{\rho^4}{\left[-\rho  \left(f' \rho '+f \rho ''\right)-f\rho '^2+1\right]^2} \\
   & +\frac{1}{4 \rho^2} \left[\rho f''+ {2 f' \rho '} \right]^2\\
   & +\frac1{4 \rho ^2} {\left[\rho f''+2 f' \rho '+4 f \rho ''\right]^2} \; .
\end{align}
It is a sum of squares, which means that all combinations in brackets
should be separately regular. Dropping the 1 and the $f\rho '^2$ in
the first bracket, since they are regular by themselves, we can take suitable linear superpositions
of such combinations to show that $f'', f' \rho'$ and $f\rho''$ should
all be regular at the horizon. 
If we further assume that the surface
gravity $\kappa = f'/2$ is finite {\it and} non-zero, we find
that $\rho'$ is regular at the horizon as well.
(The surface gravity $\kappa$ is defined by $\kappa^2 = -\sfrac12 \nabla_\mu \xi_\nu \, \nabla^\mu \xi^\nu$, 
where $\xi^\mu$ is the time-translational Killing vector, eq.~\eqref{xi}. For the metric \eqref{metric}, we get $\kappa = \sfrac12 f'$.)

The finiteness of these quantities is important for the construction
of Kruskal-like coordinates \cite{RW}. Note that Einstein's equations
are not needed.

\end{document}